\documentclass[graybox]{svmult}

\usepackage[utf8]{inputenc}
\usepackage[T1]{fontenc}

\usepackage{mathptmx}       
\usepackage{helvet}         
\usepackage{courier}        
\usepackage{type1cm}        
\usepackage{makeidx}         
\usepackage{graphicx}        
\usepackage{multicol}        
\usepackage[bottom]{footmisc}

\usepackage[final]{changes}

\usepackage{amsmath}
\usepackage{amsfonts}
\usepackage{algorithm,algorithmic}
\usepackage{subcaption}
\usepackage{multirow}

\graphicspath{{figures/}}

\algsetup{linenosize=\small}

\usepackage{hyperref}
\usepackage[english]{babel}

\makeindex             

\title*{A \replaced{44}{66}-element mesh of Schneiders' pyramid}
\subtitle{Bounding the difficulty of hex-meshing problems}
\author{Kilian Verhetsel, Jeanne Pellerin, Jean-François Remacle}

\institute{Kilian Verhetsel \at Université catholique de Louvain, Avenue Georges Lemaitre 4, bte L4.05.02, 1348 Louvain-la-Neuve, Belgium, \email{kilian.verhetsel@uclouvain.be}
 \and Jeanne Pellerin \at Université catholique de Louvain, Avenue Georges Lemaitre 4, bte L4.05.02, 1348 Louvain-la-Neuve, Belgium, \email{jeanne.pellerin@uclouvain.be}
 \and Jean-François Remacle \at Université catholique de Louvain, Avenue Georges  Lemaitre 4, bte L4.05.02, 1348 Louvain-la-Neuve, Belgium, \email{jean-francois.remacle@uclouvain.be}}

\begin{document}

\maketitle

\abstract{%
  This paper shows that constraint programming techniques can successfully be
  used to solve challenging hex-meshing problems. Schneiders' pyramid is a
  square-based pyramid whose facets are subdivided into three or four
  quadrangles by adding vertices at edge midpoints and facet centroids.  In this
  paper, we prove that Schneiders' pyramid has no hexahedral meshes with fewer
  than 18 interior vertices and 17 hexahedra, and introduce a valid mesh with
  \replaced{44}{66} hexahedra. \deleted{We also introduce a parity-changing
    operator for hexahedral meshes, simpler than the construction by Bern,
    Eppstein, and Erickson.} \added{We also construct the smallest known mesh of
    the octagonal spindle, with 40 hexahedra and 42 interior vertices.} These
  results were obtained through a general purpose algorithm that computes the
  hexahedral meshes conformal to a given quadrilateral surface boundary.  The
  lower bound \added{for Schneiders'pyramid} is obtained by exhaustively listing
  the hexahedral meshes with up to 17 interior vertices and which have the same
  boundary as the pyramid.  Our \added{44-element} mesh is obtained by
  modifying a prior solution with 88 hexahedra.  The number of elements was
  reduced \replaced{using an algorithm which locally
    simplifies}{by locally simplifying} groups of hexahedra.  Given the boundary
  of such a group, our algorithm is used to find a mesh of its interior that has
  fewer elements than the initial subdivision.  The resulting mesh is untangled
  to obtain a valid hexahedral mesh.}

\section{Introduction}

\begin{figure}
 \centering
  \includegraphics[width=0.7\textwidth]{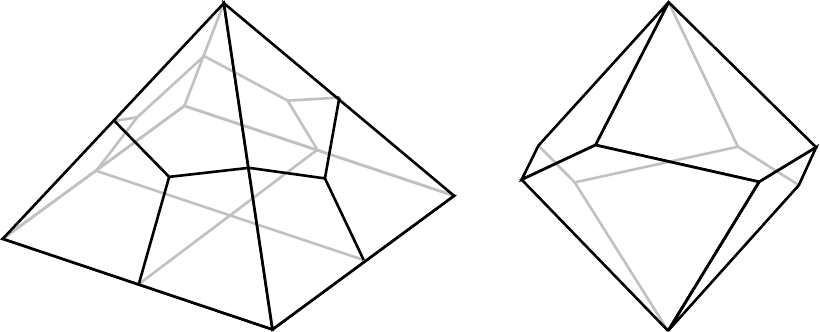}
  \caption{Left: Schneiders'pyramid. Right:  The octogonal spindle.}
  \label{fig:difficult}
\end{figure}

From the finite element practitioners point of view, hexahedral meshes have
several advantages over tetrahedral meshes.  However, there is no algorithm to
generate a hexahedral mesh conformal to a given quadrilateral boundary.  The
state-of-the-art hexahedral meshing methods fail on small polyhedra such as the
octogonal spindle \replaced{and}{or} Schneiders' pyramid (\autoref{fig:difficult}).  Schneiders'
pyramid is a square-based pyramid with eight additional vertices at edge
midpoints and five at face midpoints, and with its triangular and quadrangular
faces split into three and four quadrangles respectively.  The octogonal
spindle, or tetragonal trapezohedron, can be used to construct Schneiders'
pyramid by adding four hexahedra to form the pyramid base.  The problem of
meshing this pyramid with hexahedra was introduced by \cite{schneiders1996grid}
as an example of a boundary mesh for which no hexahedral mesh was known.

The question of the existence of a solution was settled by
\cite{mitchell1996characterization} who proved that all quadrilateral surface
meshes of the sphere with an even number of quadrilateral facets do have a
hexahedral mesh. The algorithm deduced from the proof of this important
theoretical result, as well as those of \cite{eppstein1999linear} and
\cite{erickson2014efficiently}, generates too many hexahedra to be practical.
\cite{carbonera2010constructive} constructs $5396$ times more hexahedra than
there are quadrilateral facets.  In 2002, \cite{yamakawa2002hexhoop} introduced
the hexhoop template family and constructed a hexahedral mesh of Schneiders'
pyramid with 118 hexahedra. Later on, they \deleted{used hexhoop and the geode template
  of 
  to build an 88-element solution}\added{improved their
solution, building an 88-element mesh} \cite{yamakawa2010schneiders}. \added{Very
recently, a 36-element mesh was constructed by finding a sequence of
flipping operations to transform the cube into Schneiders' pyramid, interpreting each
operation as the insertion of a hexahedron \cite{xiang36}.}

In this paper, we propose a backtracking algorithm to enumerate the
combinatorial meshes of the interior of a given quadrilateral surface
(\autoref{sec:enumerate}).  Our first contribution is to prove that there is no
hexahedral mesh of Schneiders' pyramid with strictly fewer than 12 interior
vertices.  Using the same approach, we also prove that there is no hexahedral
mesh of the octagonal spindle with strictly fewer than 21 interior vertices.  \deleted{We
answer a question raised in \cite{bern2002flipping} and show that the parity of
the number of hexahedra in a mesh can be changed by inserting only 27 hexahedra
(21 vertices).}

\begin{figure}
  \centering
  \includegraphics[width=\textwidth]{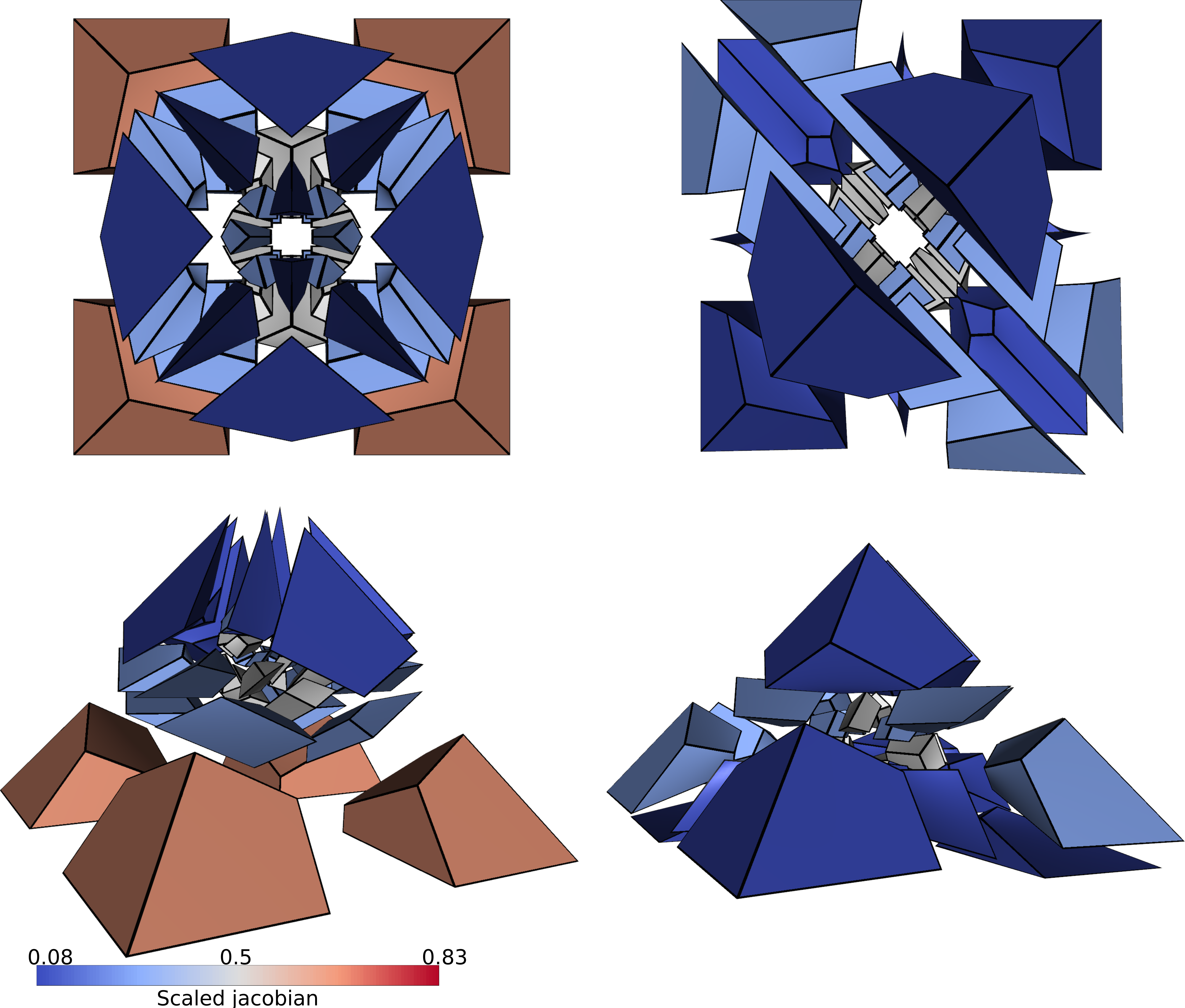}
  \caption{Comparison of our 44-element mesh of Schneiders' pyramid (left) with
    the smallest known 36-element solution (right). Both admit two planar
    symmetries.}
  \label{fig:schneiders-best}
\end{figure}

The second contribution of this paper is an algorithm \replaced{allowing the
  construction of a new}{for constructing a smaller} hexahedral mesh of
Schneiders' pyramid \replaced{and the smallest known mesh of the octagonal
  spindle}{with 66 hexahedra and 63 interior vertices}
(\autoref{fig:schneiders-best}).  This construction uses a modified version of
the backtracking algorithm to simplify the 88-element solution of
\cite{yamakawa2010schneiders} and reduce the number of hexahedra to 66 by
locally simplifying groups of hexahedra (\autoref{sec:remeshing}). The realized
operations may be viewed as a general form of cube flips
\cite{bern2002flipping}.  They substitute a set of hexahedra by another set
without changing their boundary.  However, instead of having a predefined set of
flips, as do other local operations on hexahedral meshes
\cite{tautges2003topology}, our algorithm automatically operates generically on
any group of hexahedra. This group of hexahedra is replaced by fewer hexahedra
using a combinatorial approach. The resulting mesh is untangled to obtain a
valid hexahedral mesh. \added{Furthermore, we used a sheet extraction procedure to
  construct a 40-element mesh of the octagonal spindle, and a 44-element mesh of
  the pyramid. \cite{ledoux2010topological,borden2002sheet}}

The C implementation of our algorithms and the resulting meshes can all be
downloaded from \url{https://www.hextreme.eu/}.

\section{Enumerating combinatorial hexahedral meshes}
\label{sec:enumerate}

In this section, we describe an algorithm that lists all possible hexahedral
meshes with a prescribed boundary. We use this algorithm to determine lower
bounds for the number of vertices and hexahedra needed to mesh the octagonal
spindle and Schneiders' pyramid.  It is also the key to the local mesh
simplification algorithm we propose in \autoref{sec:remeshing}.

When discussing the existence of hexahedral meshes or when enumerating those of
the interior of a given quadrilateral mesh, we first ignore geometric issues and
consider \emph{combinatorial hexahedral meshes}.  In a combinatorial hexahedral
mesh, the hexahedra are represented as sequences of 8 integers, where distinct
integers represent distinct vertices.
A set of hexahedra defines a valid combinatorial mesh if all pairs of hexahedra
are \emph{compatible}: their intersection must be a shared combinatorial face
(i.e. one of their 8 vertices, 12 edges, or 6 quadrangular facets) or be empty.
Each quadrangle is also required to either be on the boundary (i.e. in exactly
one hexahedron), or in the interior of the mesh (i.e. in exactly two hexahedra).

\subsection{Backtrack search algorithm}

Given $\partial H$, a combinatorial quad-mesh of a closed surface, $H_{max}$ a
maximum number of hexahedra, and $V_{max}$ a maximum number of vertices, our
algorithm lists all combinatorial hexahedral meshes $H$ such that:
\begin{itemize}
\item the boundary of $H$ is $\partial H$,
\item the number of hexahedra $|H|$ is at most $H_{\max}$,
\item the total number of vertices in $H$ is at most $V_{\max}$.
\end{itemize}

This problem we are solving has similarities with problems commonly encountered
in \emph{constraint programming}: (i) efficiently filtering a large set of
potential solutions and (ii) managing solutions having multiple equivalent
representations.  Our implementation adopts concepts and strategies from this
field.  For a more general study of these problems, we refer the reader to
\cite{rossi2006handbook}.

\begin{figure}
  \centering
  \includegraphics[width=0.7\textwidth]{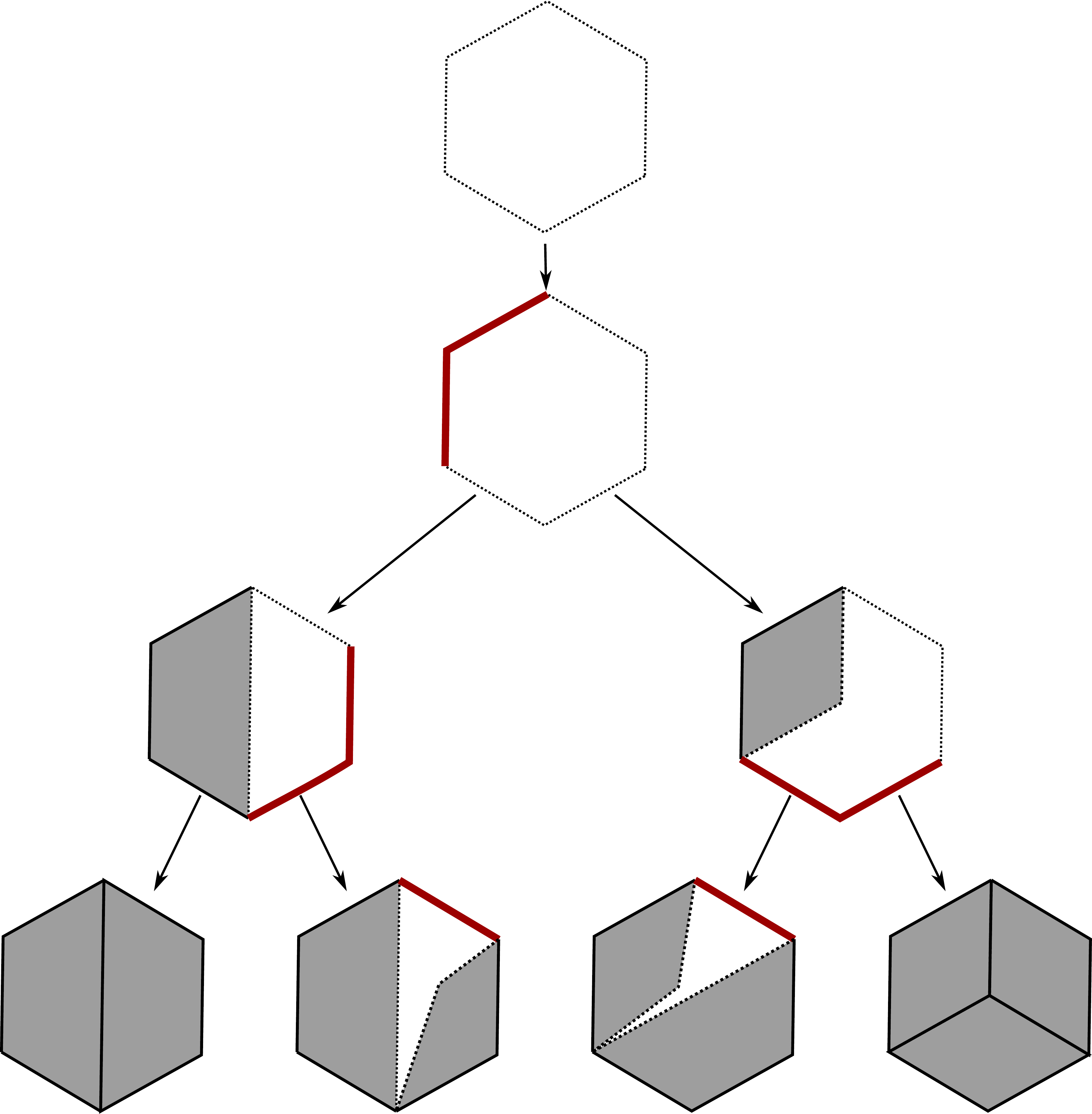}
  \caption{Searching all quadrilateral meshes of a polygon with up to one
    interior point. The search tree leaves are either valid solutions, or
    correspond to detected failure points where \autoref{alg:search} backtracks.}
  \label{fig:enumerate}
\end{figure}

The hexahedra are built one at a time by choosing a sequence of 8 vertices. At
each step, all possible candidates for one of the 8 vertices are considered and
the algorithm branches for each possibility.  Each branch corresponds to the
addition of a vertex to the current hexahedron.  When a complete solution is
determined, or when the search fails (no available candidates to complete a
hexahedron), the algorithm backtracks to the previous choice.  This process is
repeated until all possibilities have been explored.  \autoref{alg:search}
corresponds to the exploration of a search tree (\autoref{fig:enumerate}) where
each branching node represents the choice of a point, and the leaves represent
either solutions or failure points where the algorithm backtracks.  The search
tree has an exponential size in the maximum number of hexahedra in a solution.
This high complexity is managed by pruning branches that cannot contain a
solution and by using efficient implementations of all performed operations.


\begin{algorithm}
  \caption{Recursive enumeration of the hexahedral meshes of the interior of $\partial{}H$}
  \label{alg:search}
  \begin{algorithmic}[1]
    \REQUIRE $\partial{}H$, the boundary;
             $S$, a partial solution;
             $C = (C_1, \dots, C_8)$, the sets of candidate vertices for the current hexahedron
    \IF{the boundary of $S$ is $\partial H$} \label{stmt:boundary-check}
      \STATE Print solution $S$
    \ELSIF{$|S| = H_{\max}$}
      \STATE Backtrack
    \ELSE
      \STATE $C \gets \textsc{Filter-Candidates}(\partial{}H, S, C)$
      \IF{$|C_1| = \dots = |C_8| = 1$}
        \STATE $S' \gets S \cup \left\{\left(v_1, v_2, v_3, v_4, v_5, v_6, v_7, v_8\right)\right\}$
        \STATE $\textsc{Search}\left(\partial{}H, S', \textsc{Initialize-Candidates}(S')\right)$
      \ELSIF{$\min_{i \in \{1, \dots, 8\}} |C_i| = 0$}
        \STATE Backtrack
      \ELSE
        \STATE $i \gets \textsc{Pick-Hex-Vertex}(C)$ \label{stmt:pick-var}
        \FOR{each $v \in C_i$}
          \STATE $C' \gets C$
          \STATE $C'_i \gets \{v\}$
          \STATE $\textsc{Search}(\partial{}H, S, C')$
        \ENDFOR
      \ENDIF
    \ENDIF
  \end{algorithmic}
\end{algorithm}

\subsection{Search space reduction strategies}
\label{sec:filtering}

In this section, we describe the key points of our implementation of
\autoref{alg:search}, all of which aim at reducing the search space explored by
the algorithm:
\begin{itemize}
\item the order in which the hexahedra are constructed is crucial --- we use an
  advancing-front strategy and start the construction of hexahedra from the
  boundary;
\item an efficient filtering algorithm that eliminates candidate vertices that
  would create incompatible combinatorial hexahedra in the solution;
\item a method to manage the high number of symmetries of this problem;
\item the order in which the current hexahedron vertices are selected.
\end{itemize}

\begin{figure}[t]
  \centering
  \includegraphics[width=.8\textwidth]{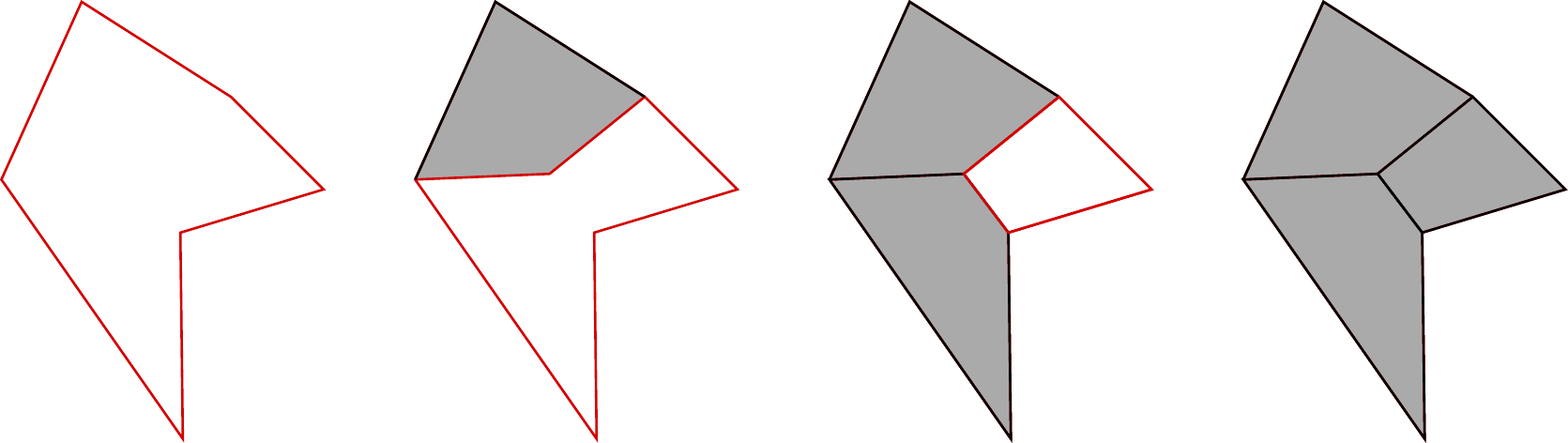}
  \caption{Each new element must share a face with the front of boundary faces
    (red).}
  \label{fig:quad-queue}
\end{figure}

\noindent \textbf{Advancing-front construction} \hspace{.3cm} While the
hexahedra of a combinatorial mesh can be arbitrarily reordered, constructing
them in a specific order makes the algorithm significantly faster. We use a
classical advancing front generation strategy and require the hexahedron under
construction to share a face with a front of quadrangles. There are then only
four vertices needed to complete a hexahedron.  The quadrangle front is
constituted of the interior facets that are in only one hexahedron, or of
boundary facets that are in no hexahedra.  At the root of the search tree, it
is set to be the boundary $\partial{}H$.  An interior facet is added to the
front after its first appearance in the mesh. The facet is removed from the front
when it is added to the partial solution. When the front becomes
empty, the boundary of the solution matches the input
(\autoref{fig:quad-queue}).

\vspace{1ex}
\noindent \textbf{Filtering out candidate vertices} \hspace{.3cm}
For each of the eight vertices of the hexahedron under construction, we store a
set of candidate vertices that could be part of the solution.  Some of these
vertices would make the current hexahedron incompatible with some already
existing hexahedra.
Therefore when initiating the construction of a hexahedron, or when adding a
vertex to a hexahedron, vertices that cannot be added without creating
incompatibilities between the current hexahedron and the already built hexahedra
are filtered out.
The following rules are used to eliminate candidates:
\begin{enumerate}
\item an edge cannot match the diagonal of an existing quadrangle, or an
  interior diagonal of an existing hexahedron;
\item conversely an interior diagonal cannot match the diagonal of an existing
  quadrangle, or an existing edge, or an interior diagonal of an existing
  hexahedron;
\item a facet diagonal cannot match an existing hexahedron edge, or an existing
  hexahedron interior diagonal;
\item if one facet diagonal matches an existing quadrangle diagonal, so must the
second one;
\item all eight vertices must be different.
\end{enumerate}

Our implementation tracks three sets of vertices for each vertex $v$. These sets
are used to build the candidate set for each vertex of a new hexahedron
(\autoref{alg:initialize-candidates}).  These sets are updated whenever an edge,
quadrangle or interior diagonal is added to the mesh, and they are:
\begin{itemize}
\item $\textsc{Allowed-Neighbors}(v)$, the set of vertices $u$ such that an edge
  $(u, v)$ could be added to the mesh without creating incompatibilities with
  existing hexahedra or facets of the boundary;
\item $\textsc{Known-Neighbors}(v)$, the set of vertices that are adjacent to
  $v$ in existing hexahedra or in quadrangles contained in the boundary;
\item $\textsc{Known-Diagonals}(v)$, the set of all vertices $u$ such that $(u,
  v)$ is one of the four interior diagonals of a hexahedron.
\end{itemize}

Because the execution time of the search algorithm blows up as the number of
vertices increases, the number of vertices each set contains is always small,
making them good candidates for being represented as bit-sets.

\begin{algorithm}[t]
  \caption{$\textsc{Initialize-candidates}(S)$: Compute the sets candidate vertices }
  \label{alg:initialize-candidates}
  \begin{algorithmic}[1]
    \REQUIRE $S$, a set of hexahedra.
    \ENSURE $C = (C_1, \dots, C_8)$, the sets of candidate vertices
    for the next hexahedron.

    \STATE Let $(v_1, v_2, v_3, v_4)$ be some quadrangle that needs to occur in the mesh.
    \FOR{each $i \in \{1, \dots, 4\}$}
      \STATE $C_i\gets \{v_i\}$
    \ENDFOR
    \FOR{each $i \in \{1, \dots, 4\}$}
      \STATE $C_{4+i} \gets \textsc{Allowed-Neighbors}(v_i) \setminus \{v_1, v_2, v_3, v_4\}$
      \FOR{each $j \in \{1, \dots, 4\}$}
        \IF{$i \ne j$}
          \STATE $C_{4+i} \gets C_{4+i} \setminus
          \textsc{Known-Diagonals}(v_j) \setminus \textsc{Known-Neighbors}(v_j)$
        \ENDIF

        \IF{$i = j + 2 \mod 4$}
          \STATE $C_{4+i} \gets C_{4+i} \setminus \{ v_k \;|\; (v_j, v_k) \mbox{ is the diagonal of a quadrangle} \}$
        \ENDIF
      \ENDFOR
    \ENDFOR
    \RETURN $(C_1, \dots, C_8)$
  \end{algorithmic}
\end{algorithm}

\begin{figure}
  \centering
  \includegraphics[width=0.7\textwidth]{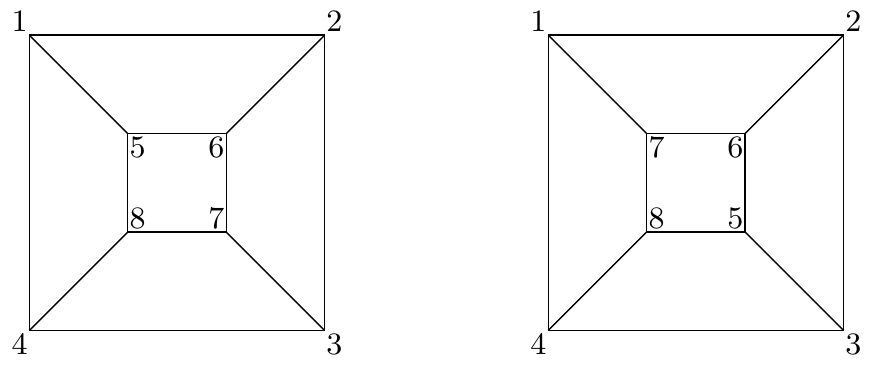}
  \caption{Two of the $4!$ ways to label the $4$ interior vertices of this
    mesh.}
  \label{fig:symmetry-interior}
\end{figure}

\begin{figure}
  \centering
  \includegraphics[width=0.6\textwidth]{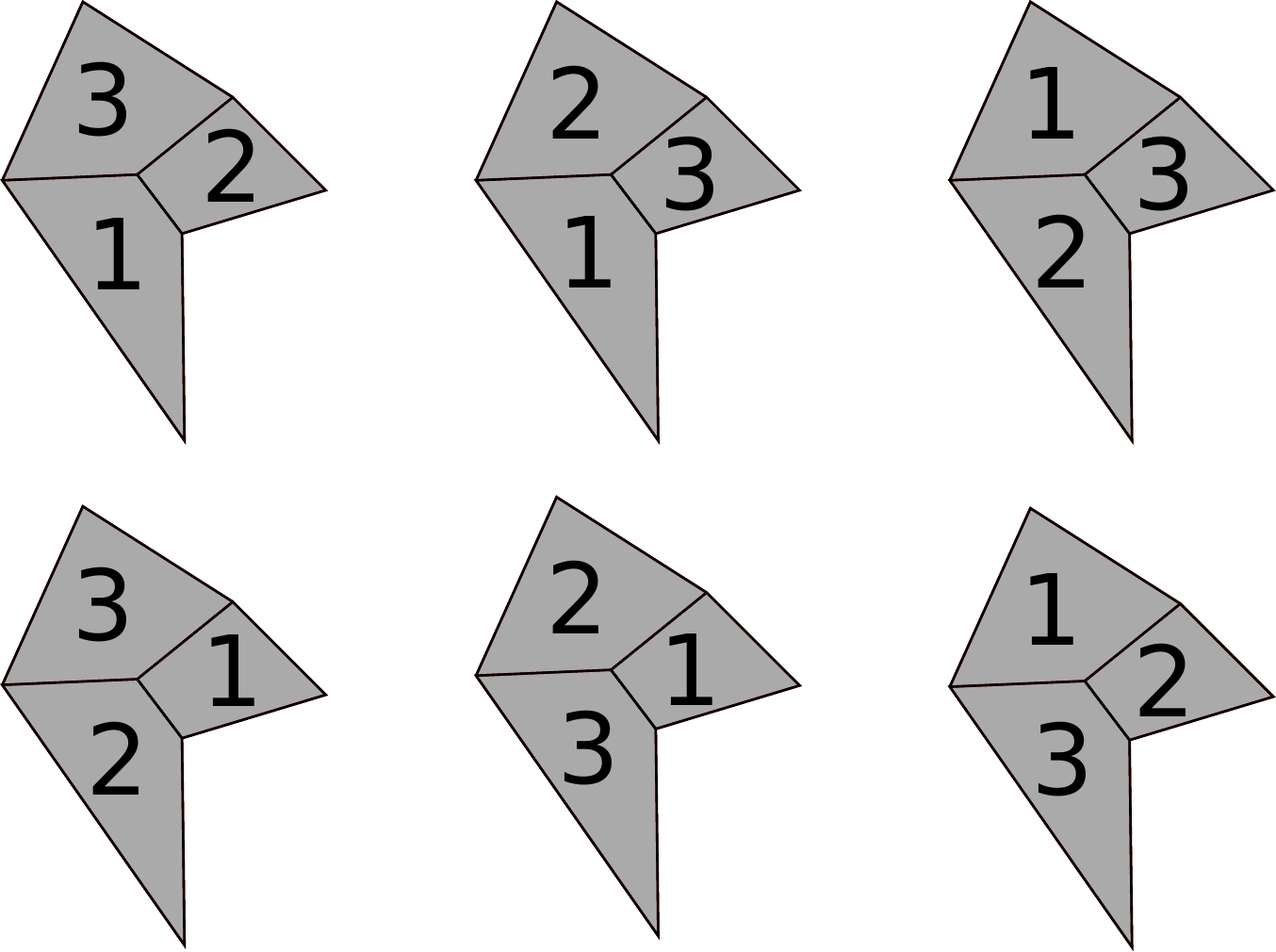}
  \caption{The $3!$ different ways to number the elements of a 3-element mesh.}
  \label{fig:permute-hex}
\end{figure}

\begin{figure}
  \centering
  \includegraphics[width=0.6\textwidth]{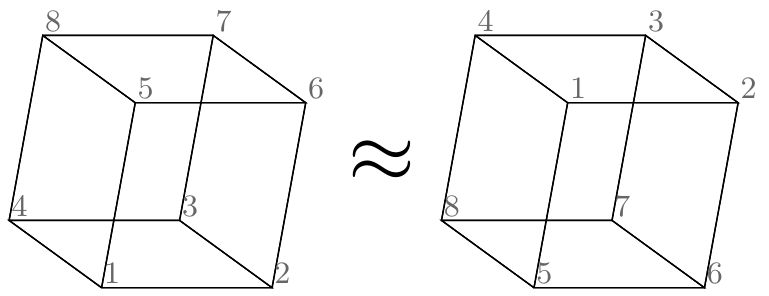}
  \caption{Two combinatorially equivalent hexahedra.}
  \label{fig:reorder-hex}
\end{figure}

\vspace{1ex}
\noindent \textbf{Symmetry breaking} \hspace{.3cm}
Combinatorial meshes are characterized by their large number of symmetries, a
major challenge when operating on combinatorial hexahedral meshes.  Indeed, a
combinatorial hexahedral mesh has many equivalent representations:
\begin{enumerate}
\item interior vertices can be relabelled (\autoref{fig:symmetry-interior}) ---
  for boundary vertices, the algorithm uses the same labels as the input;
\item the hexahedra of the solution can be constructed in a different order
  (\autoref{fig:permute-hex});
\item for a given hexahedron, written as an ordered sequence of 8 vertices,
  there are $1,680 = 8! / 24$ ways to reorder these vertices while leaving the
  hexahedron unchanged (\autoref{fig:reorder-hex}).
\end{enumerate}

The advancing front strategy defines the order in which the solution hexahedra
are constructed (symmetry 2).  This also uniquely determines the order of
vertices in a hexahedron (symmetry 3).  To prevent the relabelling of interior
vertices (symmetry 1), we add \emph{value precedence} constraints to our problem
\cite{law2004global}.  A solution $H$ found by the algorithm can be written as
an array of $8|H|$ integers, writing down the vertices of each hexahedron in the
order in which they were constructed by the algorithm.  In an array,
$x$~\emph{precedes}~$y$ when the first occurrence of~$x$ is before the first
occurrence of~$y$. Enforcing a total precedence order on interior vertices, we
guarantee that only one of their permutations is a solution.

\vspace{1ex}
\noindent \textbf{Optimization of hexahedron construction} \hspace{.3cm}
The efficiency of \autoref{alg:search} depends on the size of the search tree
needed to explore all possibilities.  A cheap approach to reduce the number of
nodes in the search tree is to choose the vertex with the smallest set of
candidate vertices when deciding which vertex to branch on
\cite{beck2004trying}.  This does not affect the correctness of the algorithm,
as long as a vertex with more than one candidate is selected.

\subsection{Parallel search}

\begin{figure}
  \centering
  \includegraphics[width=\textwidth]{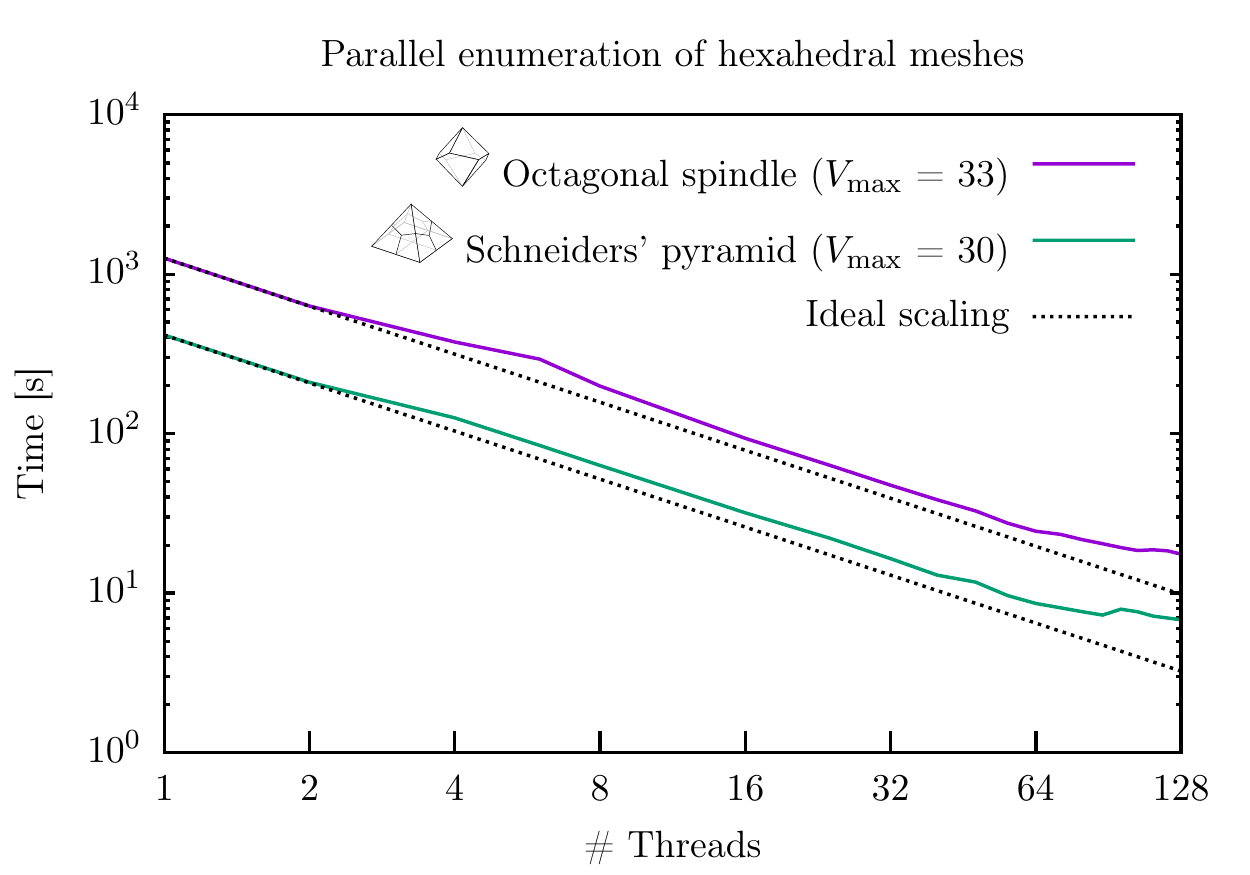}
  \caption{Time to explore a search tree in parallel on a machine with two AMD
    EPYC 7551 CPUs (32 cores each, 2 threads per core). Using 64 threads, the
    speed-up is of 48 for Schneiders' pyramid and 52 for the octagonal spindle.}
  \label{fig:parallel}
\end{figure}

The exploration a search tree can be parallelized in a natural way by exploring
different subtrees in parallel, making the algorithm much faster on parallel
architectures (\autoref{fig:parallel}). We use an approach similar to the
embarrassingly parallel search of \cite{regin2013embarrassingly}. The main
challenge to overcome is that some subtrees are multiple orders of magnitude
larger than other ones without any possibility to determine it ahead of time.

We solve this issue by attributing many subtrees to each worker thread, so that
all threads must on average perform the same amount of work (we used 4096
subtrees per thread). At the start of the search, the tree is explored in a
breadth-first manner until a layer with enough subproblems is reached. The nodes
of this layer are then explored in parallel by independent worker threads using
\autoref{alg:search}.

\begin{figure}
  \centering
  \includegraphics[width=\textwidth]{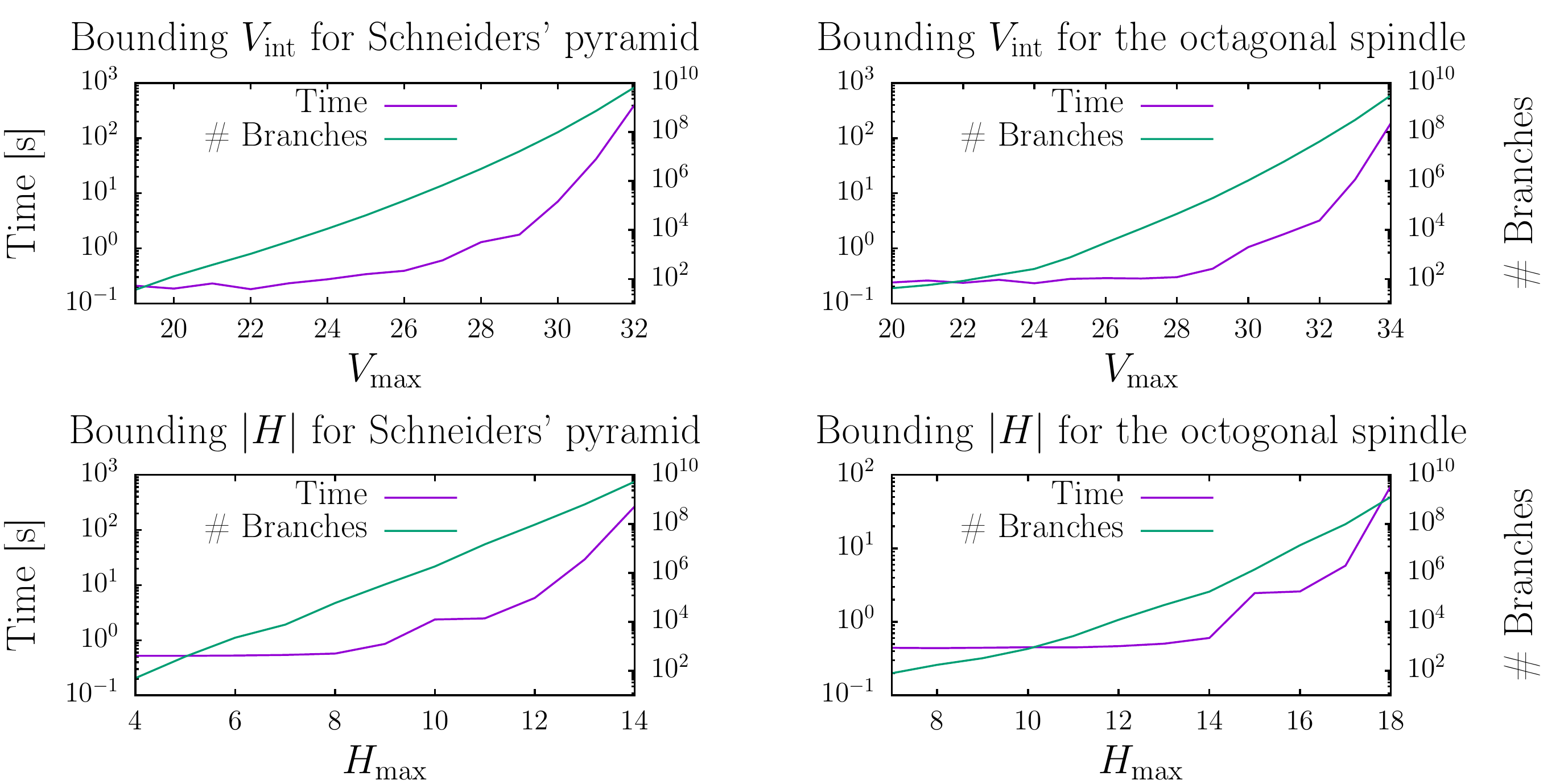}

  \caption{The time to prove lower bounds for the number of interior vertices
    $V_{\text{int}}$ and the number of hexahedra $H$ required to mesh a polyhedron
    increases exponentially. This is due to the exponential size of the search
    tree explored by the algorithm.}
  \label{fig:lower-bounds-time}
\end{figure}

\subsection{Lower bounds for hex-meshing problems}

Using \autoref{alg:search}, we computed lower bounds for the number of vertices
and hexahedra required to mesh Schneiders' pyramid and the octagonal spindle
(\autoref{fig:difficult}). The algorithm is run multiple times, and we increment
either $V_{\max}$ or $H_{\max}$ between each run. At each step, we verify that
no solution was found by the algorithm. The time required to compute these
bounds increases exponentially as the bounds become tighter
(\autoref{fig:lower-bounds-time}).

\begin{theorem}
  Any hexahedral mesh of Schneiders' pyramid has at least 18 interior vertices
  and 17 hexahedra.
\end{theorem}

\begin{theorem}
  Any hexahedral mesh of the octagonal spindle has at least 29 interior
  vertices and 21 hexahedra.
\end{theorem}





\section{Simplifying hexahedral meshes}
\label{sec:remeshing}

\begin{figure}
  \centering
  \includegraphics[width=\textwidth]{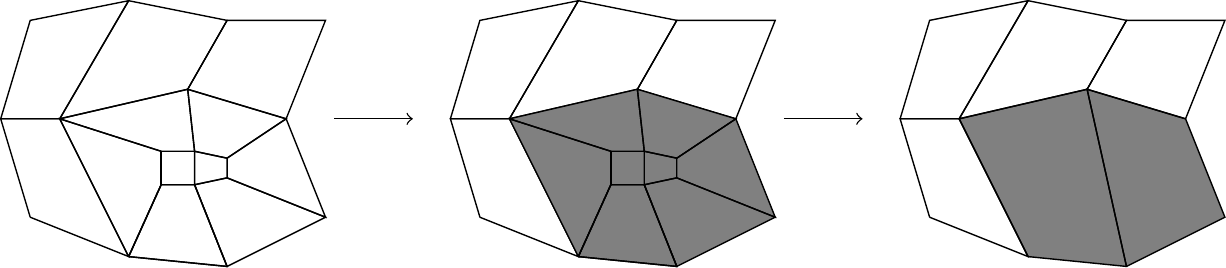}
  \caption{The number of elements in a mesh can be reduced by operating locally
    on a cavity.}
  \label{fig:remeshing}
\end{figure}

The algorithm described in the previous section can be used to find the smallest
hexahedral mesh with a given boundary. In this section, we use this algorithm to
accomplish our goal of computing upper bounds for the number of hexahedra
required to mesh Schneiders' pyramid.  From the 88-element solution of
\cite{yamakawa2010schneiders}, we locally simplify the mesh. By simplification
we mean decreasing the number of hexahedra (\autoref{fig:remeshing}).  The
realized operations may be viewed as a generalized form of cube flips
\cite{bern2002flipping} that substitute a set of hexahedra by another set
without changing their boundary.  However, instead of a finite set of
transformations, the algorithm introduced in this section automatically
determines them at execution time.

Globally minimizing the number of hexahedra in the mesh is a computationally
demanding task. Our algorithm therefore selects a small subset of the mesh, or
\emph{cavity}, and focuses on modifying the connectivity of the mesh only within
this cavity.  Our hexahedral mesh simplification algorithm is based on
\autoref{alg:search}.  From a geometric hexahedral mesh it outputs a geometric
hexahedral mesh whose boundary is strictly identical and which has fewer
elements.

The mesh simplification procedure has three main steps: 
\begin{enumerate}
\item the selection of a cavity, the group of hexahedra to simplify,
  $\mathcal{C}$;
\item finding the smallest hexahedral mesh $\mathcal{C}_{\min}$ compatible with
  the cavity boundary $\partial \mathcal{C}$ and replacing the cavity with this
  smaller mesh;
\item untangling the hexahedra to determine valid coordinates for the mesh
  vertices.
\end{enumerate}

\noindent \textbf{1. Cavity selection} \hspace{.3cm}
The cavity selection algorithm is a greedy algorithm that starts from a random
element of the input hexahedral mesh (\autoref{alg:cavity}).  When the target
size, in terms of number of hexahedra is reached, this process stops.  The
choice of a target cavity size is a trade-off between the cost of finding the
hexahedral meshes of the cavity and the likelihood that the mesh can be
simplified by remeshing the cavity. Cavities with many hexahedra are more likely
to accept smaller meshes, but the cost of finding the smallest hexahedral mesh
$\mathcal{C}_{\min}$ increases exponentially with the number of hexahedra in the
cavity. In practice, we start by considering relatively small cavities
containing up to 10 hexahedra, and increase this limit when no improvement is
possible.  We require the cavity to contain at least 4 interior
vertices. Indeed, when there are no interior vertices (e.g. with a stack of
hexahedra), it is not possible to remove any hexahedra. As the number of
interior vertices increases, so does the likelihood that the cavity can be
simplified.

\begin{algorithm}
  \caption{Cavity selection algorithm}
  \label{alg:cavity}
  \begin{algorithmic}[1]
    \REQUIRE $H$, the mesh; $n$, the size of the cavity
    \ENSURE A cavity $\mathcal{C}$ of $n$ elements
    \STATE $h \gets $ a random element of $H$
    \STATE $\mathcal{C} \gets \{h\}$
    \WHILE{$|\mathcal{C}| \ne n$}
      \STATE $h \gets $ a random element of $H \setminus \mathcal{C}$ sharing a facet
      with a hexahedron in $\mathcal{C}$
      \STATE $\mathcal{C} \gets \mathcal{C} \cup \{h\}$
    \ENDWHILE
    \RETURN $\mathcal{C}$
  \end{algorithmic}
\end{algorithm}

\vspace{1ex}
\textbf{2. Cavity remeshing}\hspace{.3cm}
To find a smaller mesh of the boundary of a cavity $\mathcal{C}$, we first solve
the combinatorial problem, i.e. we find the smallest combinatorial hexahedral
mesh of $\partial \mathcal{C}$, and then solve the geometric problem of finding
valid coordinates for the modified mesh vertices.

The combinatorial problem of finding the smallest mesh of $\partial \mathcal{C}$
is but an application of \autoref{alg:search} which enumerates all combinatorial
meshes of a given surface.  The maximum number of hexahedra $H_{max}$ of the
solution is set to a smaller value than $|\mathcal{C}|$.  Changing the parity of
a hexahedral mesh is known to be a difficult operation
\cite{schwartz2004construction}, so we set $H_{max}$ to $|\mathcal{C}| - 2$.  We
also set the limit to the number of interior vertices $V_{max}$ to one less than
the number of interior vertices in $\mathcal{C}$ to accelerate the search.

\begin{figure}[t]
  \centering
  \includegraphics[width=0.5\textwidth]{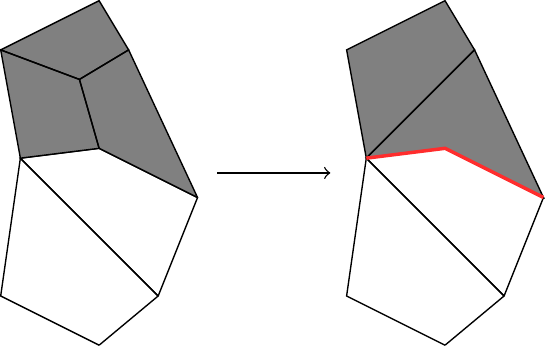}
  \caption{Replacing the cavity with a valid mesh sharing the same boundary
    still produces an invalid mesh by creating two quadrangles sharing two edges.}
  \label{fig:remeshing-incompatibilities}
\end{figure}

There is a subtle but important difference between meshing a cavity in an
existing mesh and meshing a stand-alone polyhedron: the hexahedra inside the
cavity must be compatible with the other elements of the input mesh.  An example
where new elements from a cavity are not compatible with elements adjacent to
the cavity is given in \autoref{fig:remeshing-incompatibilities}.  A 3-element
cavity is replaced by 2 quadrangles, but one of these two quadrangles shares two
edges with an existing element, which is an invalid configuration.  To guarantee
that the algorithm does not break the mesh validity, the data structures used to
filter out inadequate vertex candidates (\autoref{sec:filtering}) are modified
to take into account the hexahedra that are not part of the cavity.
 
\begin{figure}[t]
  \centering
  \includegraphics[width=0.8\textwidth]{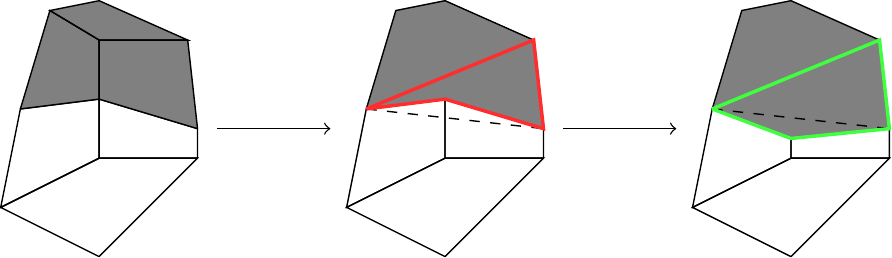}
  \caption{A valid change to the connectivity of the mesh can create a
    geometrically invalid mesh, fixed by moving the vertices.}
  \label{fig:remeshing-invalidities}
\end{figure}

\vspace{1ex}
\textbf{3. Untangling} \hspace{.3cm}
The previous step of the algorithm found a new connectivity for the mesh.  The
simplified mesh obtained by using this result is not valid in general because
the interiors of hexahedra may intersect (\autoref{fig:remeshing-invalidities}).
To obtain a valid geometric mesh we use the untangling algorithm described in
\cite{toulorge2013robust}.  The vertices are iteratively moved until all
hexahedra in the mesh are valid.  If the untangling fails, connectivity changes
are undone.  The validity of the final mesh is evaluated with the method
proposed by \cite{johnen2017validity}.

\vspace{1ex}
\noindent\textbf{A 66-element mesh of Schneiders' pyramid} \hspace{.3cm}

\begin{table}[b]
  \centering
  \begin{tabular}{cc c ccc c cc c cc}
    \multicolumn{2}{c}{Initial mesh} && \multicolumn{3}{c}{Initial cavity} & & \multicolumn{2}{c}{Remeshed cavity} & & \multicolumn{2}{c}{New mesh} \\
     \#hex & \#vert. & & \#hex & \#vert. & \#bd. facets & & \#hex & \#vert.& & \#hex & \#vert. \\
    \hline
    88 & 105 &  & 8  & 23 & 18 &  & 6  & 21 &  & 86 & 103 \\
    86 & 103 &  & 8  & 23 & 18 &  & 6  & 21 &  & 84 & 101 \\
    84 & 101 &  & 8  & 23 & 18 &  & 6  & 21 &  & 82 & 99  \\
    82 & 99  &  & 14 & 33 & 24 &  & 8  & 27 &  & 76 & 93  \\
    76 & 93  &  & 6  & 16 & 10 &  & 2  & 12 &  & 72 & 89  \\
    72 & 89  &  & 18 & 40 & 30 &  & 12 & 32 &  & 66 & 81  \\
  \end{tabular}
  \caption{Cavity remeshing operations performed by our hex-mesh simplification
    algorithm on Yamakawa's 88-element mesh of Schneiders' pyramid
    \cite{yamakawa2010schneiders}.}
  \label{table:iterations}
\end{table}

\begin{figure}
  \centering
  \includegraphics[width=0.4\textwidth]{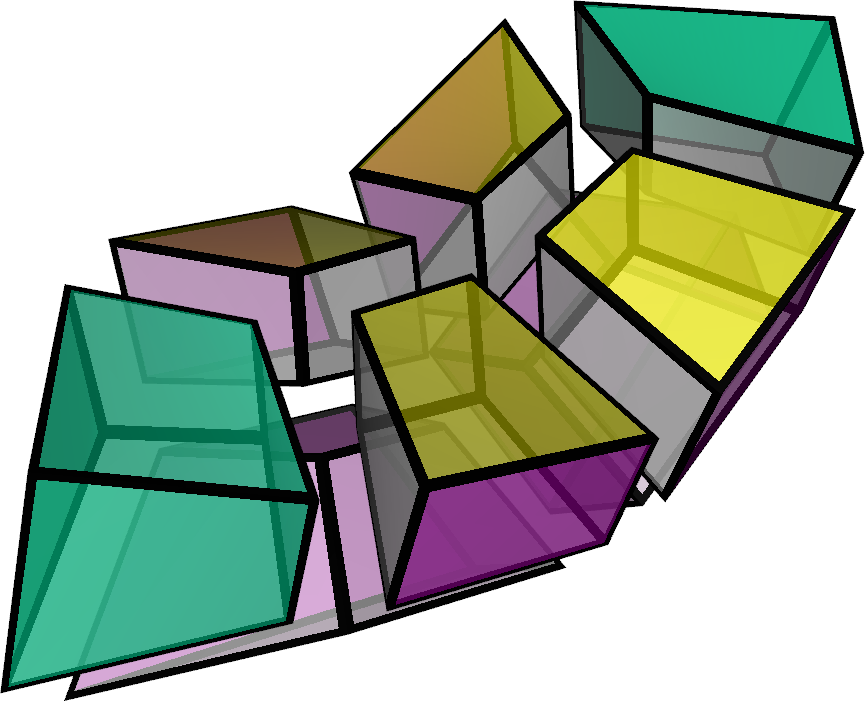}~\hfill~%
  \includegraphics[width=0.4\textwidth]{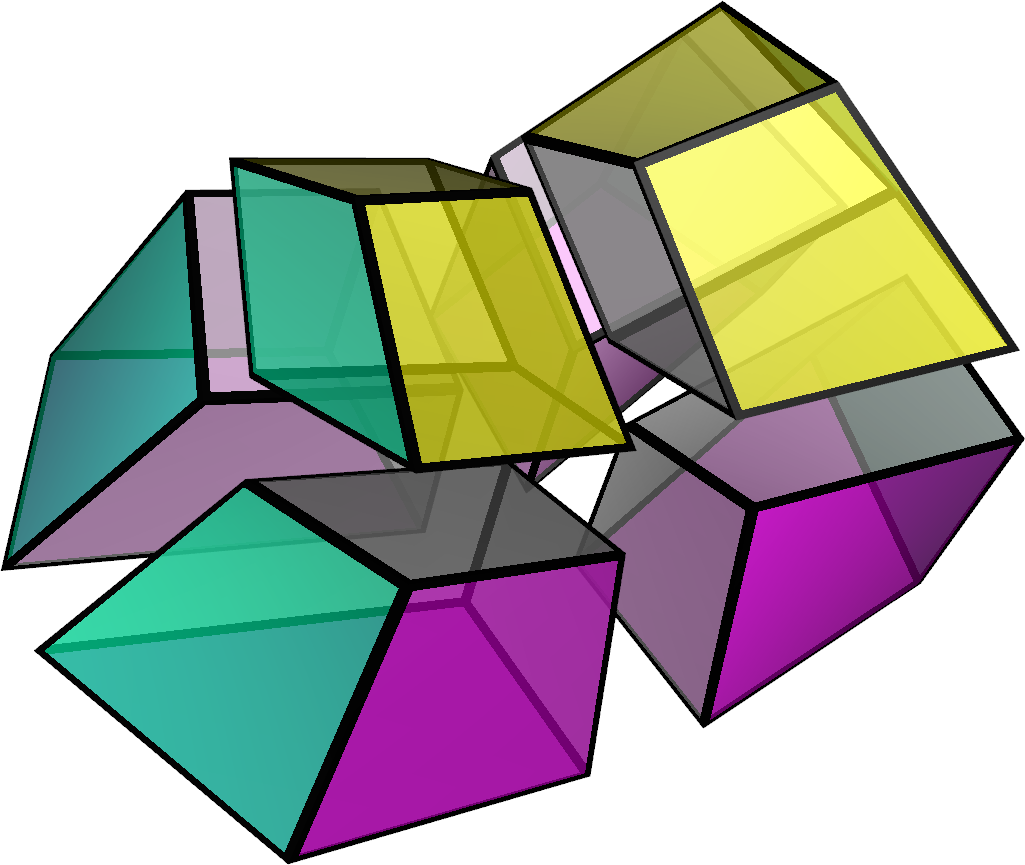}
  \includegraphics[width=0.4\textwidth]{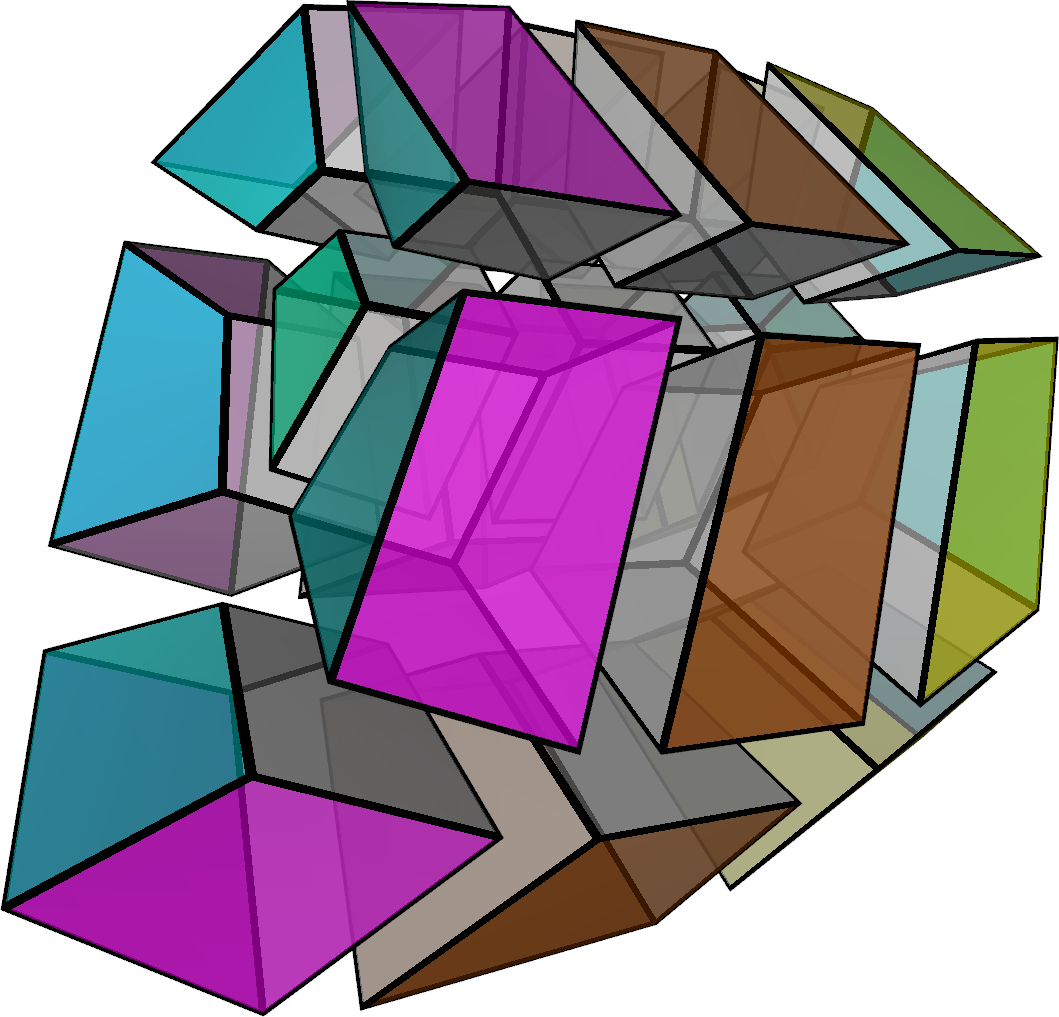}~\hfill~%
  \includegraphics[width=0.4\textwidth]{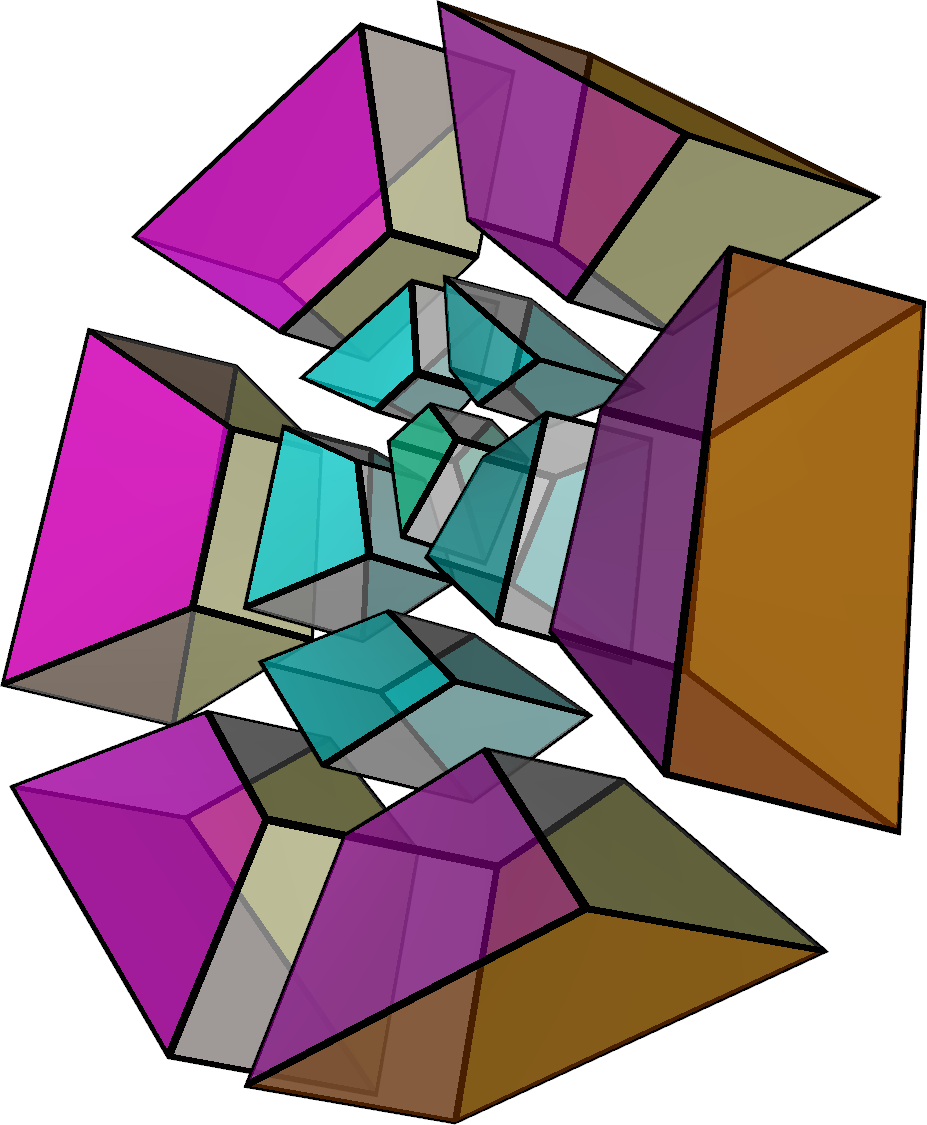}
  \caption{(top) Removal of two hexahedra from Schneiders' pyramid; (bottom)
    removal of six hexahedra. The initial cavity (left) and the remeshed cavity
    (right) have the same combinatorial boundary (top: 18 facets; bottom: 30
    facets). Colors highlight the correspondence between faces.}
  \label{fig:pretty-cavity}
\end{figure}

We applied our algorithm to Yamakawa's mesh of Schneiders' pyramid and obtained
a valid hexahedral mesh with 66 hexahedra and 63 interior vertices
(\autoref{fig:schneiders-best}).  \autoref{table:iterations} shows the sizes of
the different cavities simplified by our algorithm. It takes a few minutes for
our algorithm to reduce the number of hexahedra in the mesh from 88 down to 66
in our final mesh. \autoref{fig:pretty-cavity} shows the changes to the
connectivity of the mesh performed in two different iterations of the
algorithm. The vertices had to be moved to obtain a valid mesh, but the
combinatorial boundary remains the same. For example, for the second pair of
cavities in the figure, the same 30 facets can be seen before and after the
remeshing operation: there is a central facet, surrounded by a ring of five
quadrangles, followed by three rings of six quadrangles, followed by one more
ring of five quadrangles surrounding a single face.  We also determined a
combinatorial mesh with 64 hexahedra and 59 interior vertices, on which the
untangling failed.

\vspace{1ex}
\noindent\textbf{A 44-element mesh of Schneiders' pyramid} \hspace{.3cm}

We used the 72-element mesh described in the previous paragraph to create a
40-element mesh of the octagonal spindle. Indeed, such a mesh can be constructed
by extracting one of the two sheets of the dual of our 72-element solution using
the method described in \cite{borden2002sheet}. The mesh resulting from this
operation is the smallest known mesh of the octagonal spindle, with 40 hexahedra
and 42 interior vertices. Our 44-element mesh of the pyramid mesh was obtained
by adding 4 hexahedra to our mesh of the spindle
(\autoref{fig:schneiders-best}).

All the meshes discussed in this section are available online at
\url{https://www.hextreme.eu/}.

\section{Conclusion}

The main contribution of this paper is an algorithm to prove new bounds for
Schneiders' pyramid and the octagonal spindle hex-meshing problems. The
relatively large number of vertices required to mesh the pyramid implies that
subdividing pyramids into hexahedra to create all-hexahedral meshes will
necessarily create many additional hexahedra. This makes it likely that some of
the hexahedra created in this manner will be invalid. Because these algorithms
are general, they could be used to evaluate the viability of other subdivision
schemes to create all-hexahedral meshes.

One limitation of the hex-mesh simplification algorithm described in this paper
is that its execution becomes expensive, because finding the smallest hexahedral
mesh of a cavity becomes exponentially more time consuming as its size
increases. A cheaper algorithm could be designed by finding a small set of local
operations and an algorithm to choose which of them to perform in order to
reduce the size of the mesh.

\begin{acknowledgement}
  This research is supported by the European Research Council (project HEXTREME,
  ERC-2015-AdG-694020). Computational resources have been provided by the
  supercomputing facilities of the Université catholique de Louvain (CISM/UCL)
  and the Consortium des Équipements de Calcul Intensif en Fédération Wallonie
  Bruxelles (CÉCI) funded by the Fond de la Recherche Scientifique de Belgique
  (F.R.S.-FNRS) under convention 2.5020.11.
\end{acknowledgement}

\bibliographystyle{plain}
\bibliography{doc}

\begin{thebibliography}{10}

\bibitem{beck2004trying}
J~Christopher Beck, Patrick Prosser, and Richard~J Wallace.
\newblock Trying again to fail-first.
\newblock In {\em International Workshop on Constraint Solving and Constraint
  Logic Programming}, pages 41--55. Springer, 2004.

\bibitem{bern2002flipping}
Marshall Bern, David Eppstein, and Jeff Erickson.
\newblock Flipping cubical meshes.
\newblock {\em Engineering with Computers}, 18(3):173--187, 2002.

\bibitem{borden2002sheet}
Michael~J Borden, Steven~E Benzley, and Jason~F Shepherd.
\newblock Hexahedral sheet extraction.
\newblock In {\em IMR}, pages 147--152, 2002.

\bibitem{carbonera2010constructive}
Carlos~D Carbonera and Jason~F Shepherd.
\newblock A constructive approach to constrained hexahedral mesh generation.
\newblock {\em Engineering with Computers}, 26(4):341--350, 2010.

\bibitem{eppstein1999linear}
David Eppstein.
\newblock Linear complexity hexahedral mesh generation.
\newblock {\em Computational Geometry}, 12(1-2):3--16, 1999.

\bibitem{erickson2014efficiently}
Jeff Erickson.
\newblock Efficiently hex-meshing things with topology.
\newblock {\em Discrete \& Computational Geometry}, 52(3):427--449, 2014.

\bibitem{johnen2017validity}
Amaury Johnen, J-C Weill, and J-F Remacle.
\newblock Robust and efficient validation of the linear hexahedral element.
\newblock {\em Procedia Engineering}, 203:271--283, 2017.

\bibitem{law2004global}
Yat~Chiu Law and Jimmy~HM Lee.
\newblock Global constraints for integer and set value precedence.
\newblock In {\em International Conference on Principles and Practice of
  Constraint Programming}, pages 362--376. Springer, 2004.

\bibitem{ledoux2010topological}
Franck Ledoux and Jason Shepherd.
\newblock Topological modifications of hexahedral meshes via sheet operations:
  a theoretical study.
\newblock {\em Engineering with Computers}, 26(4):433--447, 2010.

\bibitem{mitchell1996characterization}
Scott~A Mitchell.
\newblock A characterization of the quadrilateral meshes of a surface which
  admit a compatible hexahedral mesh of the enclosed volume.
\newblock In {\em Annual Symposium on Theoretical Aspects of Computer Science},
  pages 465--476. Springer, 1996.

\bibitem{regin2013embarrassingly}
Jean-Charles R{\'e}gin, Mohamed Rezgui, and Arnaud Malapert.
\newblock Embarrassingly parallel search.
\newblock In {\em International Conference on Principles and Practice of
  Constraint Programming}, pages 596--610. Springer, 2013.

\bibitem{rossi2006handbook}
Francesca Rossi, Peter Van~Beek, and Toby Walsh.
\newblock {\em Handbook of constraint programming}.
\newblock Elsevier, 2006.

\bibitem{schneiders1996grid}
Robert Schneiders.
\newblock A grid-based algorithm for the generation of hexahedral element
  meshes.
\newblock {\em Engineering with computers}, 12(3-4):168--177, 1996.

\bibitem{schwartz2004construction}
Alexander Schwartz and G{\"u}nter~M Ziegler.
\newblock Construction techniques for cubical complexes, odd cubical
  4-polytopes, and prescribed dual manifolds.
\newblock {\em Experimental Mathematics}, 13(4):385--413, 2004.

\bibitem{tautges2003topology}
Timothy~J Tautges and Sarah~E Knoop.
\newblock Topology modification of hexahedral meshes using atomic dual-based
  operations.
\newblock {\em algorithms}, 11:12, 2003.

\bibitem{toulorge2013robust}
Thomas Toulorge, Christophe Geuzaine, Jean-Fran{\c{c}}ois Remacle, and Jonathan
  Lambrechts.
\newblock Robust untangling of curvilinear meshes.
\newblock {\em Journal of Computational Physics}, 254:8--26, 2013.

\bibitem{xiang36}
Shang Xiang and Jianfei Liu.
\newblock A 36-element solution to schneiders' pyramid hex-meshing problem and
  a parity-changing template for hex-mesh revision.
\newblock {\em arXiv preprint arXiv:1807.09415}, 2018.

\bibitem{yamakawa2002hexhoop}
Soji Yamakawa and Kenji Shimada.
\newblock Hexhoop: modular templates for converting a hex-dominant mesh to an
  all-hex mesh.
\newblock {\em Engineering with computers}, 18(3):211--228, 2002.

\bibitem{yamakawa2010schneiders}
Soji Yamakawa and Kenji Shimada.
\newblock 88-element solution to schneiders' pyramid hex-meshing problem.
\newblock {\em International Journal for Numerical Methods in Biomedical
  Engineering}, 26(12):1700--1712, 2010.

\end{thebibliography}

\end{document}